\documentclass[aps,pre,10pt,twocolumn,superscriptaddress]{revtex4-2}
\usepackage{graphicx}
\usepackage{epstopdf}
\usepackage{xcolor}
\usepackage{amsmath}
\usepackage{amssymb}
\usepackage[utf8]{inputenc}
\usepackage{graphicx,float,hyperref}
\usepackage{natbib}
\usepackage{graphicx,latexsym} 

\begin{document}


\title{Performance analysis of quantum harmonic Otto engine and refrigerator under a trade-off figure of merit}

%
\author{Kirandeep Kaur}
\email{kiran.nitj@gmail.com}
\affiliation{ Department of Physical Sciences,   
Indian Institute of Science Education and Research Mohali,
Sector 81, S.A.S. Nagar, Manauli PO 140306, Punjab, India}

\author{Shishram Rebari}
\email{rebaris@nitj.ac.in}
\affiliation{ Department of Physics,   Dr B R Ambedkar National Institute of Technology Jalandhar, Punjab-144011, India}

\author{Varinder Singh}
\email{varinder@ibs.re.kr}
\affiliation{Center for Theoretical Physics of Complex Systems, Institute for Basic Science (IBS), Daejeon 34126, Korea}
%
%


\begin{abstract}
We investigate the optimal performance of  quantum Otto engine and refrigeration cycles of a time-dependent harmonic oscillator under a trade-off figure of merit for both adiabatic and nonadiabatic (sudden-switch) frequency modulations. For heat engine (refrigerator), the chosen trade-off figure of merit is an objective function defined by the product of efficiency (coefficient of performance) and work output (cooling load), thus  representing a compromise between them. We obtain analytical expressions for the efficiency and coefficient of performance of the harmonic Otto cycle for the optimal performance of the thermal machine in various operational regimes. Particularly, in the sudden-switch regime, we discuss the implications of the nonadiabatic driving on the performance of the thermal machine under consideration, and obtain analytic expressions for the maximum  achievable efficiency and coefficient of performance  of the harmonic Otto thermal machine. Further, by carrying out a detailed comparative  analysis  of the heat engine operating under the chosen trade-off objective function with one operating at maximum work output, we show that the trade-off objective functions have desirable operation only for the adiabatic driving whereas for the sudden switch operation, the choice of a trade-off objective function does not make much difference as the performance of the engine is dominated by frictional effects. 
\end{abstract}

\pacs{03.67.Lx, 03.67.Bg}

\maketitle 

\section{Introduction}
Thermal machines have played very important role in the development of classical theory of thermodynamics. The discovery of Carnot efficiency, a universal efficiency which puts an  upper bound on the efficiency of all thermal machines working between two thermal reservoirs at different temperature, led Clausius to formulate the second law of thermodynamics. Heat engines and refrigerators are two primary examples of thermal machines.   A heat engine is a device which converts thermal energy into useful work by utilizing the temperature difference between two  reservoirs whereas in the case of a refrigerator, work is invested to extract heat from a cold reservoir and dump it to a hot reservoir. For thermal heat engines (refrigerators), the ideal Carnot efficiency (coefficient of performance (COP)) is given by a simple formula $\eta_C=1-\beta_h/\beta_c$ ($\zeta_C=\beta_h/(\beta_c-\beta_h)$), where $\beta_h$ and $\beta_c$ are inverse temperatures of the hot and cold reservoirs. Unfortunately, the Carnot efficiency can only be obtained under the reversible conditions, which in turn results in vanishing  power output of the engine. However, the primary goal of a practical heat engine is to generate finite power in finite-time \cite{CA1975,Chen2001}. Finite-time thermodynamics addresses this question and provides us with tools to study the optimal performance of energy conversion devices under more realistic conditions and constraints \cite{Salamon2001,Chen2001,Andresen2011,Berry1984,devosbook}. 

The common approach to investigate the optimal performance of heat engines is optimization of the power output of the engine \cite{CA1975,Esposito2010,Rezek2006,Rezek2017,GevaKosloff1992,Dorfman2018,Apertet2012B,WangHeWu2012,Lutz2012,SchmiedlSeifert2008,VandenBroeck2005,Esposito2010PRE,Tu2008,Schmiedl2007EPL,Esposito}. However, with the rising concerns over the impact of fossil fuels on the environment, the practical heat engines  operating  at maximum  power do not provide us with good choice as they dump a large amount of  entropy to the environment ultimately polluting it along with wasting a lot of fuel \cite{Andresen2011,AB1991,Hernandez2001,VJ2019}. Thus, ecologically friendly and cost-effective engines should  operate under the  of optimal conditions of such trade-off figure of merit (or objective  function) which pays equal attention to both power and efficiency (or entropy production) of the engine \cite{Andresen2011,AB1991,devosbook}. Efficient power function is such an alternative trade-off objective function which  is defined by the product of the efficiency and power of the engine. It was introduced by Stucki while studying biochemical energy conversion process \cite{Stucki}.  Yan-Chen extended this idea to study the optimal performance of endoreversible heat engines \cite{YanChen1996}. Afterwards, efficient power function has been applied to analyze performance of various classical \cite{Yilmaz,VJ2018,Zhang2017}, mesoscopic \cite{Me5}, and quantum heat engines \cite{MePRR,Deffner2020}.  Efficient power function is also shown to be well suited objective function to   study the energy conversion process in thermionic generators \cite{ChenDing},  steady and non-steady electric circuits \cite{Valencia2017}  and biological systems \cite{Stucki,Chimal,ABrown2008}.

Unlike heat engines, choosing a suitable objective function for the refrigerators is not a straight-forward task \cite{YanChen1990,Abah2016EPL,Apertet2013A}. The ideal figure of merit would be the cooling power of the refrigerator. However, for many models of refrigerators, cooling power is not a good choice as the maximization of cooling power yields the vanishing COP, which is not a useful result \cite{VJ2020,SinghSingh2022,Apertet2013A}. Also from the ecological point of view, the refrigerator should operate in a regime which pays equal attention to both COP and cooling power of the refrigerator \cite{Chen2001,YanChen1989,Kiran2021}.  The $\chi$-function, introduced by Yan and Chen \cite{YanChen1990}, is the most popular choice  of trade-off objective function  to study the optimal performance of refrigerators. As it is defined by the product of the COP and the cooling load, it automatically takes care of the trade-off between them. 

Numerous studies have been conducted to analyze the performance of classical thermal machines in finite-time thermodynamics. But recent efforts toward the miniaturization has pushed the applicability of thermal machines at the quantum level.  
For example, integration of an atomic scale refrigerator to quantum computational devices  can make them more compact, which can be used for widespread access of quantum technologies \cite{Sai2016,Sourav2021}. Further, the rapid development in the field of quantum technologies has bring up the question of resource consumption in the thermodynamic landscape \cite{Auffeves2111}. Quantum thermal devices provide the natural platform for addressing the  limits of energy consumption at the quantum level.   

In this work, we study the optimal performance of a quantum Otto engine (refrigerator) \cite{Quan2007,Kieu2004,Rezek2006,Lutz2012,VOzgur2020,Bijay2021,Barish2021,Vahid2021,Assis2019,Assis2020A,Tanmoy2021,Ozgur2017,SinghSingh2022,Pritam2023} whose working fluid is a  driven harmonic oscillator. The choice of the model is motivated by its recent experimental realizations \cite{Rossnagel2016Science,Klaers2017}. Further, the amenability of the model to analytic results makes it more suitable  to work with. For investigating optimal performance of heat engine, we choose a trade-off objective function which is defined by the product of efficiency and work output of the engine. For the lack of a better name, we will call it ``efficient work function" analogous to so called efficient power function.
The optimal performance of the refrigerator is analyzed by choosing $\chi$-criterion as the objective function. 

Th paper is organized as follows. In Sec. II we present the model of a harmonic Otto cycle consists of a harmonic oscillator as the working fluid and  two thermal reservoirs at different temperatures. In Sec. III we present analytic results for the efficiency at maximum efficient work function for both adiabatic and sudden-switch driving protocols in high- and low temperature limits. In Sec. IV the same analysis as done for the heat engine case is repeated for the harmonic Otto refrigerator. We conclude in Sec. V.

\section{Quantum Otto cycle}
The operation of a quantum Otto cycle consists  four stages: two  isochoric and two adiabatic (in the thermodynamic sense) stages \cite{Lutz2012,Abah2016EPL}.  In the following, we will briefly describe the all four stages of  a harmonic quantum Otto cycle.  
(1) Adiabatic compression $A\longrightarrow B$:  Initially,  the system is assumed to be thermalized at an 
inverse temperature $\beta_c$ (or temperature $T_c$). Then, the working medium (harmonic oscillator here)  is isolated from the environment and the frequency ($\omega$) of the quantum oscillator is changed from an initial value $\omega_c$ to a final value $\omega_h$ via an external unitary protocol. In this step, the average energy of the system increases. Because of the unitary nature of the driving protocol,  the von Neumann entropy of the oscillator remains invariant during the driving process.
(2) Hot isochore $B\longrightarrow C$: During the hot isochoric stage, the frequency of the harmonic oscillator is kept at a constant value $\omega_h$ and it is allowed to exchange heat with the hot reservoir  at  inverse temperature $\beta_h$. If the heat exchange between the system and the reservoir takes place for sufficiently long time,  the system reaches the thermal state  at inverse temperature  $\beta_h$.
(3) Isentropic expansion $C \longrightarrow D$: Again,  the system is isolated from the reservoirs and the frequency of the oscillator is unitarily changed  back from $\omega_h$ to its initial value $\omega_c$. As the average energy of the harmonic oscillator decreases in this stage, the work is done by the system.  
(4) Cold isochore $D\longrightarrow A$: Finally, the system is brought  in contact with the cold reservoir at inverse temperature $\beta_c$ ($\beta_c>\beta_h$) and frequency of the oscillator is kept at a constant value  $\omega_c$. By exchanging heat with the cold reservoir , the system relaxes back to its initial thermal state $A$.

 The mean  energies (denoted by $\langle H \rangle$) of the oscillator at   four different stages of the Otto cycle are \cite{Lutz2012,Abah2016EPL} 

\begin{equation}
\langle H\rangle_A =\frac{ \omega_c}{2}\text{coth}\Big(\frac{\beta_c  \omega_c}{2}\Big) ,
\end{equation}
\begin{equation}
\langle H\rangle_B =\frac{ \omega_h}{2} \lambda \text{coth}\Big(\frac{\beta_c  \omega_c}{2}\Big) ,
\end{equation}
\begin{equation}
\langle H\rangle_C =\frac{ \omega_h}{2} \text{coth}\Big(\frac{\beta_h   \omega_h}{2}\Big) ,
\end{equation}
\begin{equation}
\langle H\rangle_D =\frac{ \omega_c}{2}\lambda \text{coth}\Big(\frac{\beta_h  \omega_h}{2}\Big) ,
\end{equation}
where we have set $k_B=\hbar=1$. $\lambda$ is the dimensionless adiabaticity parameter which depends on the speed of the adiabatic evolution   \cite{Deffner2008,Husimi}. In general, we have $\lambda \ge 1$, and the general form of $\lambda$ is given by {Deffner2008,Husimi}
 \begin{equation}
\lambda=\frac{1}{2\omega_c\omega_h}\Big\{\omega_c^2\,\big[\omega_h^2\,X(t)^2+\dot X(t)^2\big]+\big[\omega_h^2\,Y(t)^2+\dot Y(t)^2\big]\Big\},
\nonumber
\end{equation}
 where $X(t)$ and $Y(t)$ are the solutions of the equation, $d^2X/dt^2+\omega^2(t)X=0$, satisfying $X(0)=0$, $\dot{X}(0)=1$, $Y(0)=1$, 
 $\dot{Y}(0)=0$ \cite{Deffner2008,Husimi}. 

The expression for the average heat exchanged during the hot isochore and the  cold isochore are given by the following equations
 \begin{eqnarray}
   Q_h   &=& \langle H \rangle_C-\langle H \rangle_B \nonumber
  \\
  &=& \frac{ \omega_h}{2}\Big[\text{coth}\Big(\frac{\beta_h \omega_h}{2}\Big)-\lambda\text{coth}\Big(\frac{\beta_c  \omega_c}{2}\Big) \Big]  \label{heat2}
 \end{eqnarray}
  \begin{eqnarray}
  Q_c   &=& \langle H \rangle_A-\langle H \rangle_D \nonumber
  \\
  &=& \frac{  \omega_c}{2}\Big[\text{coth}\Big(\frac{\beta_c \omega_c}{2}\Big)-\lambda\text{coth}\Big(\frac{\beta_h\omega_h}{2}\Big) \Big], \label{heat4}   
 \end{eqnarray}
where we have used a sign convention in which all the incoming fluxes (heat and work) entering into the the system are taken to be positive.
  \begin{figure}
 \centering
 \includegraphics[scale=0.65,keepaspectratio=true]{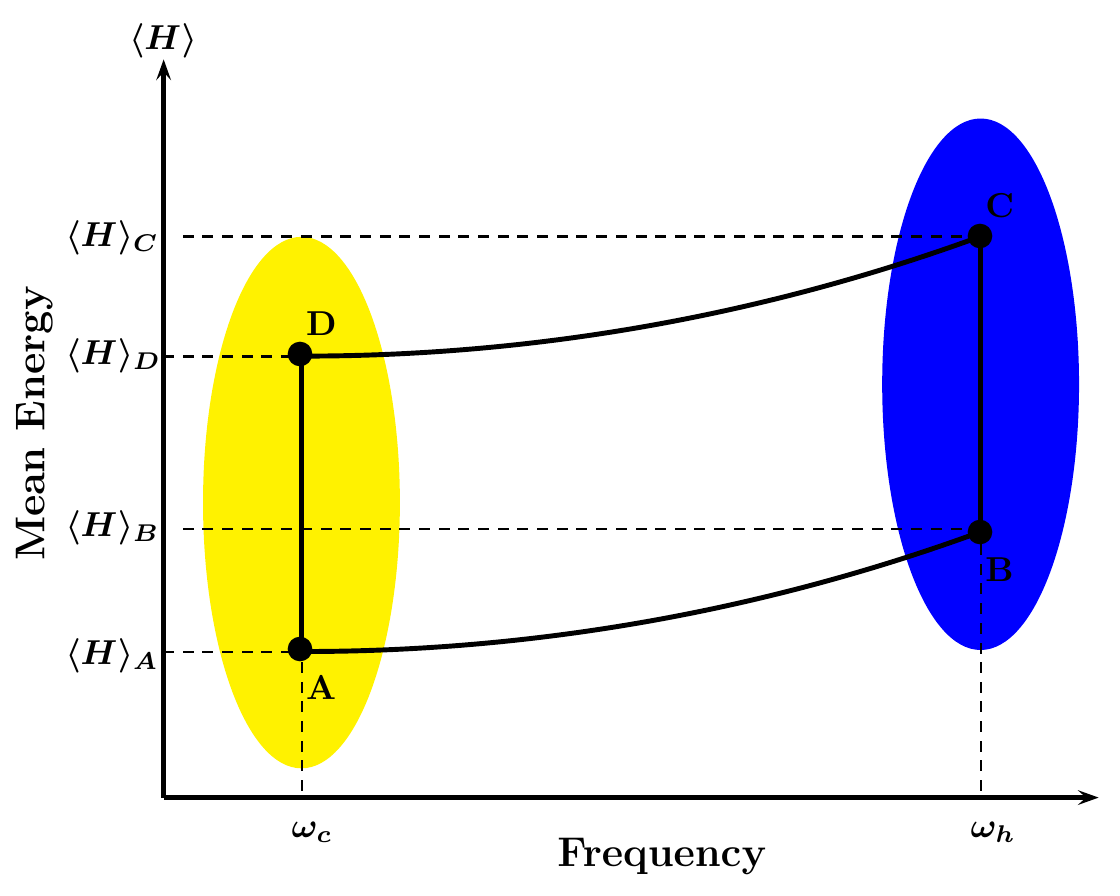}
 \caption{Pictorial depiction of  Otto cycle. The thermodynamic cycle consists of four stages: two adiabatic (A$\rightarrow$ B and C$\rightarrow$ D) and two isochoric (B$\rightarrow$ C and D $\rightarrow$ A) steps.} \label{fig:cycle}
\end{figure}
\section{Quantum Otto heat engine}
Using the first law of thermodynamics,   the net work done on the system in a complete cycle is given by, $W=-(Q_h+Q_c)$. Work is said to be extracted in one complete cycle when $W_{\rm ext}=-W=Q_h+Q_c>0$.  The general expression for the efficiency of the engine is given by
\begin{equation}
\eta = \frac{W_{\rm ext}}{Q_h} 
=
1-\frac{\omega_c}{\omega_h}\frac{\text{coth}(\beta_c   \omega_c/2)-\lambda\text{coth}(\beta_h\omega_h /2)}{\lambda \text{coth}(\beta_c  \omega_c/2)-\text{coth}(\beta_h  \omega_h /2)}. \label{effi}
\end{equation}
In this work, we 
will discuss two extreme cases: the adiabatic and sudden switch of frequencies.  For the  adiabatic process, $\lambda=1$ and for the sudden switch of frequencies,   $\lambda=(\omega_c^2+\omega_h^2)/2 \omega_c \omega_h$ \cite{{Lutz2012,Abah2016EPL}}.
\subsection{Adiabatic case}
We   start our discussion with the adiabatic case first. If the unitary evolution taking place during the adiabatic stages of the cycle is   much slower than the typical time scales of the system, we can apply quantum adiabatic theorem in our discussion. In this case, $\lambda=1$. Using Eqs.(\ref{effi}),  we have the following expression  for the efficiency of the engine 
\begin{equation}
 \eta = 1-\frac{\omega_c}{\omega_h}  \equiv 1-z, \label{effadiabatic}
 \end{equation}
which depends only on the ratio of oscillator frequencies.

\subsubsection{High-temperature regime}
In order to obtain analytic expression for the efficiency, we will work in the high temperature regime. For high temperatures,   we can approximate  $\coth(\beta_i\omega_i/2)\approx 2/(\beta_i\omega_i)$ ($i=c, h$).  Using Eqs. (\ref{heat2}) and (\ref{heat4})   in  the expression $W_{\rm ext}=Q_c+Q_h$, we have
 \begin{equation}
W_{\rm ext}=\frac{1-z}{\beta_h}+\frac{z-1}{\beta_c z}. \label{WHT}
\end{equation}
The efficient work function is given by product of efficiency [Eq. (\ref{effadiabatic})] and extracted work $W_{\rm ext}$ given in Eq. (\ref{WHT}),
\begin{equation}
W_{\eta }=  \eta \, W_{\rm ext} = (1-z) \left(\frac{1-z}{\beta_h}+\frac{z-1}{\beta_c z}\right), \label{EPHT}
\end{equation}
where $z\equiv \omega_c/\omega_h$ is the compression ratio of the Otto cycle, and $\tau=\beta_h/\beta_c=1-\eta _{\rm C}$. 

Optimizing   Eq. (\ref{EPHT})  with respect to  the compression ratio $z$  and substituting the resulting expression for $z$ in Eq. (\ref{effadiabatic}), we obtain following expression for the efficiency at maximum efficient work,
\begin{equation}
\eta^{\rm EW}_{\rm HT}=         1-\frac{1}{4}(1-\eta_C) \left ( 1+\sqrt {1 + \frac{8}{1-\eta_C}}   \right), \label{EMEP}
\end{equation}
where the superscript EW on $\eta$ in the left hand side of Eq. (\ref{EMEP}) represent  the optimization of efficient work function and subscript HT stands for high-temperature limit. Similarly, we will use superscript W to represent optimization of work output and LT will stand for low-temperature. The same expression for the efficiency can also be obtained for the optimization of endoreversible \cite{Yilmaz,YanChen1996} and symmetric low-dissipation \cite{VJ2018} models of classical heat engines. The result obtained above is not surprising as the high-temperature limit is considered to be  classical limit and   quantum heat engines are expected to behave like classical ones in this limit \cite{Geva1992,Rezek2006,VJ2019,VJ2020} .

The corresponding efficiency for  the harmonic Otto engine at maximum work  output is given by square-root formula of  Curzon and Ahlborn \cite{CA1975}, $\eta^{\rm W}_{\rm HT}=\eta_{\rm CA}=1-\sqrt{1-\eta_{\rm C}}$ \cite{Lutz2012,Rezek2006}.  In order to compare the performance of the engine operating in the maximum efficient work regime to the engine at maximum  work, we expand $\eta^{\rm EW}_{\rm HT}$ and  $\eta^{\rm W}_{\rm HT}$ in Taylor's series up to the third order term in $\eta_C$:
\begin{eqnarray}
\eta^{\rm EW}_{\rm HT} &=& \frac{2\eta_{C}}{3} + \frac{2\eta_{C}^2}{27} +    \frac{10\eta_{C}^3}{243}  + O(\eta_C^4), \label{TaylorHT}
\\
\eta^{\rm W}_{\rm HT} &=& \frac{\eta_{C}}{2} + \frac{\eta_{C}^2}{8} +    \frac{\eta_{C}^3}{16}  + O(\eta_C^4). \label{TaylorCA}
\end{eqnarray}
Clearly,  $\eta^{\rm EW}_{\rm HT} $ is always greater than $\eta^{\rm W}_{\rm HT} $, which is expected outcome \cite{Chen2001,VJ2018} as optimization of the efficient work function takes care of both efficiency and work output of the engine. For further comparison of  $\eta^{\rm EW}_{\rm HT}$ (dashed red curve)  and $\eta^{\rm W}_{\rm HT}$ (dot-dashed orange curve), we have plotted them in Fig. 2.
\subsubsection{Low-temperature regime}
Here, we investigate the performance analysis of the harmonic Otto engine in the low-temperature regime which is the favorable regime for the experimental realization of quantum heat engines as decoherence effects are minimal in this regime.   We assume that $\beta_i\omega_i\gg 1$, and set $\coth(\beta_{i}\omega_{i}/2)\approx 1+2e^{-\beta_{i}\omega_{i}}$ ($i=c, h$). Using Eqs. (\ref{heat2}) and  (\ref{heat4}) in the expression, $W_{\rm ext}=Q_h+Q_c$, and multiplying the resulting expression with Eq. (\ref{effadiabatic}), we obtain the following expression for the efficient work function
\begin{equation}
	W_{\eta }=\frac{(\omega_h-\omega_c)^2}{\omega_h}  \left(e^{-\beta_h \omega_h}-e^{-\beta_c \omega_c}\right) \label{EPLT}
\end{equation}
Unlike the case of high-temperature limit (see Eq. (\ref{EPHT})), the efficient work function of the engine cannot be written in terms of the compression ratio $z=\omega_c/\omega_h$. Hence, in order to obtain the analytic expression for the efficiency at maximum efficient work function in terms of system parameters only, we will perform a two-parameter optimization with respect to the  control parameters $\omega_h$ and $\omega_c$. 
Setting $\partial W_\eta/\partial\omega_h=0$ and $\partial W_\eta/\partial\omega_c=0$, and after a little simplification, we   obtain the following two equations, respectively:

\begin{eqnarray}
e^{\beta_h \omega_h-\beta_c \omega_c}&=&1-\frac{\beta_h \left(\omega_h-\omega_c\right) \omega_h}{\omega_h+\omega_c}, \label{A1}
\\
e^{\beta_h \omega_h-\beta_c \omega_c}&=&\frac{2}{2+\beta_c \left(\omega_h-\omega_c\right)}. \label{A2}
\end{eqnarray}
We cannot solve these two equations to obtain the analytic expressions for $\omega_h$ and $\omega_c$. However, combining Eqs. (\ref{effadiabatic}), (\ref{A1}) and (\ref{A2}), and writing in terms of $\eta_C=1-\beta_h/\beta_c$, the following transcendental equation can be obtained 
 \begin{eqnarray}
 \frac{\left(2 \eta _C-\eta \right) \left(\eta -\eta _C\right)}{\eta  \left(1-\eta _C\right)}=\ln \left[\frac{2 \left(1-\eta _C\right)}{2-\eta }\right]. \label{solLT}
 \end{eqnarray}
It is very interesting to note that the efficiency in Eq. (\ref{solLT}) does not depend on the system-parameters and depends on $\eta _C$ (or ratio of reservoir temperatures) only.  Eq. (\ref{solLT}) (solid blue curve labelled by  $\eta^{\rm EW}_{\rm LT}$) is plotted in Fig. 2. Using the information that efficiency depends on $\eta_C$ only, we can find a perturbative solution of Eq.  (\ref{solLT})  in term of $\eta_C$ for near equilibrium conditions. 
 By substituting $\eta= a_0\eta_C + a_1 \eta_C^2 + a_2 \eta_C^3 + O(\eta_C^4)$ in Eq.  (\ref{solLT})  and expanding the resulting equation in $\eta_C$. The coefficiencts of $a_0$, $a_1$ and  $a_2$ can be found recursively by solving order by order in $\eta_C$.  The first, second and third order terms are given by  $a_0 = 2/3$, $a_1=2/27$ and $a_2 = 11/243$, respectively. So near equilibrium, efficiency at maximum efficient work function behaves as follows
\begin{equation}
\eta^{\rm}_{\rm LT} = \frac{2\eta_{C}}{3} + \frac{2\eta_{C}^2}{27} +    \frac{11\eta_{C}^3}{243}  + O(\eta_C^4). \label{TaylorLT}
\end{equation} 
 We note that the first two terms in Eq. (\ref{TaylorLT}) are same as in Eq. (\ref{TaylorHT}), and the third term is slightly different from each other.  Many models of heat engines show this kind of universality in the first two terms of the efficiency at maximum power \cite{Esposito2009,Tu2008} and efficiency at maximum efficient power \cite{Zhang2017,VJ2018,MePRR}. For heat engines operating under the conditions of maximum efficient power, obeying the tight-coupling condition (no heat leaks), universality of   the first two terms is proven by Zhang and coauthor  by using the framework of stochastic thermodynamics  \cite{Zhang2017}. 
 
 For the completeness sake, we also present here the efficiency at maximum   work output of the engine operating in the low-temperature (LT) regime \cite{VSOO2021}:
 \begin{equation}
 \eta^{W}_{\rm LT} = \frac{\eta_C^2}{\eta_C-(1-\eta_C)\ln(1-\eta_C)} = \frac{\eta_C}{2}+\frac{\eta_C^2}{8}+\frac{7\eta_C^3}{96} + O(\eta_C^4), \label{TaylorFR}
 \end{equation}
 which again confirms the presence of the first two universal terms $\eta_C/2$ and $\eta_C^2/8$.  Again, it is clear from Eqs. (\ref{TaylorLT}) and (\ref{TaylorFR}) that the engine operating at maximum efficient work is more efficient as compared to the engine at maximum work output. In Fig. 2, dotted brown curve represents  Eq.   (\ref{TaylorFR}). Further, comparing Eqs. (\ref{TaylorHT}) and (\ref{TaylorLT}), it is evident that engines operating in low-temperature regime is slightly more efficient than their counterparts operating in the high-temperature regime. The same conclusion can also be drawn by comparing Eqs. (\ref{TaylorCA}) and (\ref{TaylorFR}).
\subsection{Sudden switch of frequencies}
Next, we discuss the case in which the harmonic oscillator is driven nonadiabatically at finite speed during the adiabatic branches (in thermodynamic sense). In order to derive the analytic results for the efficiency at optimal performance, we consider a special case of nonadiabatic evolution: the sudden switch protocol. In this case, the adiabaticity parameter  $\lambda$ is given by $\lambda=(\omega_c^2+\omega_h^2)/2 \omega_c \omega_h$ \cite{Deffner2008}. In terms of Otto cycle compression ratio $z=\omega_c/\omega_h$, $\lambda=(z^2+1)/2z$. The expressions for  absorbed heat $Q_h$ and extracted work $W_{\rm ext}$ take the following forms:
 \begin{equation}
 Q^{\rm SS}_h = 1 - \frac{\tau (z^2  + 1)}{2z^2}, \quad W^{\rm SS}_{\rm ext} =   \frac{(1-z^2)(z^2-\tau)} {2z^2 }. \label{Wss}
 \end{equation}
 The efficiency is no longer given by the simple formula given in Eq. (\ref{effadiabatic}), and expression for the efficiency, $\eta_{\rm SS}=W^{\rm SS}_{\rm ext}/ Q^{\rm SS}_h $, in the high-temperature limit, reads as
\begin{equation}
\eta_{\rm SS} = \frac{\left(1-z^2 \right) \left(z^2-\tau \right)}{ (2-\tau) z^2-\tau}. \label{effss} 
\end{equation}
The efficient work function $W^{\rm SS}_{\eta}=W\times\eta$ is simply obtained by multiplying $\eta_{\rm SS} $ and $W^{\rm SS}_{\rm ext} $,
\begin{equation}
W^{\rm SS}_\eta =  \frac{\left(1-z^2 \right)^2 \left(z^2-\tau \right)^2}{2z^2[ (2-\tau) z^2-\tau]}. \label{petaSS}
\end{equation}
Before discussing the optimization of the efficient work function, we highlight a couple of important points about the  engine operating in nonadiabatic regime.  First,  we can see that the positive work condition ($W^{\rm SS}_{\rm ext}>0$) in Eq. (\ref{Wss}) lead to the constrain $z^2>\tau$ on the operation of the  engine. This puts more stringent condition on the work extraction for a nonadiabatic harmonic Otto engine as compared to its adiabatic counterpart in which positive working condition is simply given by $z>\tau$. Thus, for the given ratio of the temperatures of the cold and hot reservoirs, the compression ratio $z$ for the sudden switch (in general for any nonadiabatic driving protocol) case should be large as compared as its adiabatic counterpart to extract positive work from the engine cycle. Second important point is that unlike the adiabatic case, the maximum efficiency of the engine in the nonadiabatic regime is no longer given by Carnot efficiency due to its highly frictional nature \cite{VOzgur2020,Plastina2014}. The origin of friction in the cycle due to sudden switch protocol can be explained as below. During the adiabatic (in the thermodynamic sense), the sudden quench of the frequency of the quantum harmonic oscillator induces the transitions among the different energy levels of the oscillator, thus generating coherences in between them.  Generation of the coherence (off-diagonal elements of the system density matrix in instantaneous energy eigenbasis of the Hamiltonian) results in extra energetic cost as compared to adiabatic driving protocol, and an additional parasitic internal energy is stored in the system (harmonic oscillator). During the proceeding isochoric stages of the cycle, this extra cost gets dissipated to the heat baths  and is termed as quantum friction \cite{Rezek2010,Feldmann2006,Ozgur2017,Plastina2014}.

\begin{figure}   
 \begin{center}
\includegraphics[width=8.6cm]{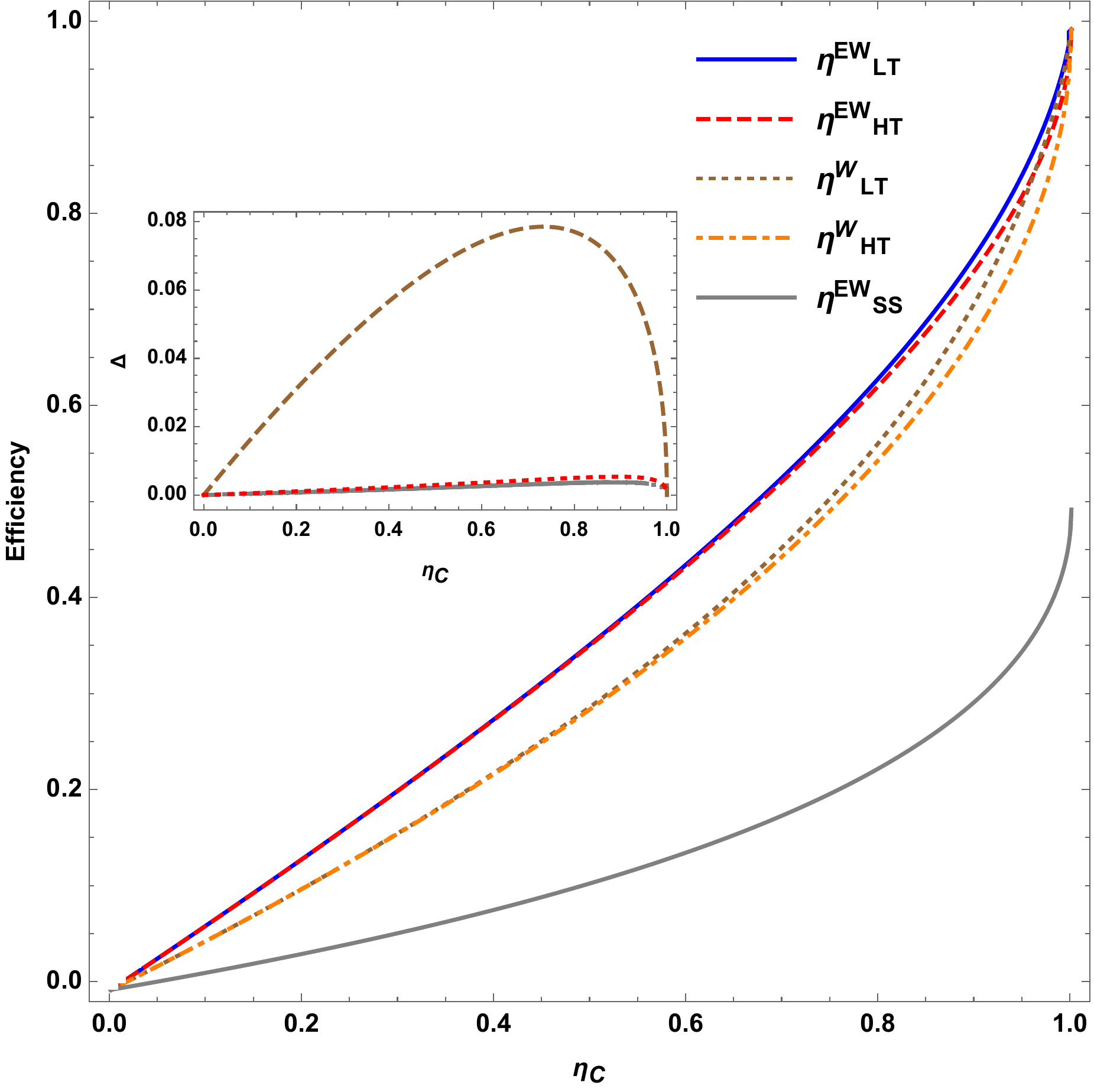}
 \end{center}
\caption{Efficiency at maximum efficient work function versus Carnot efficiency $\eta_C$. For the adiabatic case, solid blue curve represents the low-temperature limit, Eq. (\ref{solLT}), while the dashed red curve represents  the high-temperature limit, Eq. (\ref{EMEP}). Dotted brown   and dot-dashed orange curves represent the corresponding efficiencies at maximum  work output in the low and high-temperature limits, respectively. Lower lying gray curve represents efficiency at maximum efficient work function for the sudden switch regime in the high-temperature limit, Eq. (\ref{OESS}). In the inset, we have plotted the difference ($\Delta$) between efficiency at maximum efficient work function  and  and efficiency at maximum work for both adiabatic (dashed brown curve) and sudden-switch (solid gray curve) cases in the high temperature limit. Dotted red curve in the inset represent difference, $\Delta=
\eta^{\rm SS}_{\rm max}-\eta^{\rm W}_{\rm SS}$, between the maximum possible efficiency and efficiency at maximum work.}
\end{figure}
\subsubsection{Maximum efficiency in the sudden switch regime}

For the special case of sudden switch protocol, we find the expression for the maximum efficiency of the harmonic Otto engine.  By optimizing Eq. (\ref{effss}) with respect to the compression ratio $z$, and substituting the resulting expression for $z$ back in Eq.  (\ref{effss}), we obtain the following expression for the maximum efficiency in the sudden switch case
\begin{equation}
\eta^{\rm SS}_{\rm max} = \frac{\left(3-\eta_C-2 \sqrt{2 (1-\eta_C)}\right) \eta_C}{(1+\eta_C)^2}. \label{etamaxss}
\end{equation}
The same expression for the maximum efficiency of the harmonic Otto engine woking in the sudden switch regime was also derived in Ref. \cite{VOzgur2020} using a different approach. It was also noted that the $\eta^{\rm SS}_{\rm max}$ is even smaller than the half the Carnot efficiency, i. e., $\eta^{\rm SS}_{\rm max} \leq \eta_C/2$.

Although, in order to obtain the analytic expression (Eq. (\ref{etamaxss})) for the  maximum efficiency, we have worked in the high-temperature regime, we have numerically checked that  the result holds for general case. 
To prove it, we plot a histogram, for the given temperatures, of the sampled values of the efficiency given in Eq. (\ref{etamaxss}) for randomly sampling over a region of the parametric space ($\omega_c,\,\omega_h$).  For chosen values of $\beta_c=1$ and $\beta_h=1/12$, $\eta^{\rm SS}_{\rm max} =0.32$. We can clearly see that in Fig. 3, all the sampled values of efficiency $\eta^{\rm SS}$ lies below this bound, which confirms our assertion that $\eta^{\rm SS}_{\rm max}$ serves as an upper bound in all operational regimes. While uniformly sampling the parametric space ($\omega_c,\,\omega_h$), we choose $\omega_{c, h}\in [0,30]$ so that the parametric space spans all operational regimes not just the high-temperature regime.

\subsubsection{Efficiency at maximum efficient work function}
Now, we turn our attention to the optimization of the efficient work function with respect to the compression ratio  $z$. Setting $\partial P^{\rm SS}_\eta/\partial z=0$, and solving for   $z$, we obtain
\begin{equation}
z^{*}=\sqrt{\frac{5 \tau ^2-2 \tau ^3-4 \tau -\tau  A -A^2}{2 (\tau -2) A}} , \label{optimalzz}
\end{equation}
$A=\sqrt[3]{(1-\tau)(2-\tau) \sqrt{(13-8 \tau ) \tau -16} \tau ^{3/2}+\tau ^4-2 \tau ^2}$.
Substituting Eq. (\ref{optimalzz}) in Eq. (\ref{effss}) and (\ref{petaSS}), we obtain the following expressions for efficiency at maximum efficient work function,
\begin{equation}
\eta^{\rm EW}_{\rm SS} = -\frac{\left(A^2+A \tau  (2 \tau -3)+B \tau \right) \left(A^2+A (3 \tau -4)+B \tau \right)}{2 A (\tau -2)^2 \left(A^2-A \tau +B \tau \right)}, \label{OESS}
\end{equation}
%
%
%
%
where $B=\tau  (2 \tau -5)+4$.  We plot Eq. (\ref{OESS}) in Fig. 2 (see solid gray curve). The corresponding efficiency at maximum work is given by 
$\eta^{\rm W}_{\rm SS}=(1-\sqrt{1-\eta_C})/(2+\sqrt{1-\eta_C})$ \cite{Rezek2006}. By plotting the difference  between them in inset of Fig. 2 (see solid gray curve in the inset), we notice that both the efficiencies lies very close to each other.  The dotted red curve in the inset of Fig. 2, representing the difference $\Delta=\eta^{\rm SS}_{\rm max}-\eta^{\rm W}_{\rm SS}$, indicates that maximum power and maximum efficiency points lie very close to each other. 

%
 %

\begin{figure}   
 \begin{center}
\includegraphics[width=8.6cm]{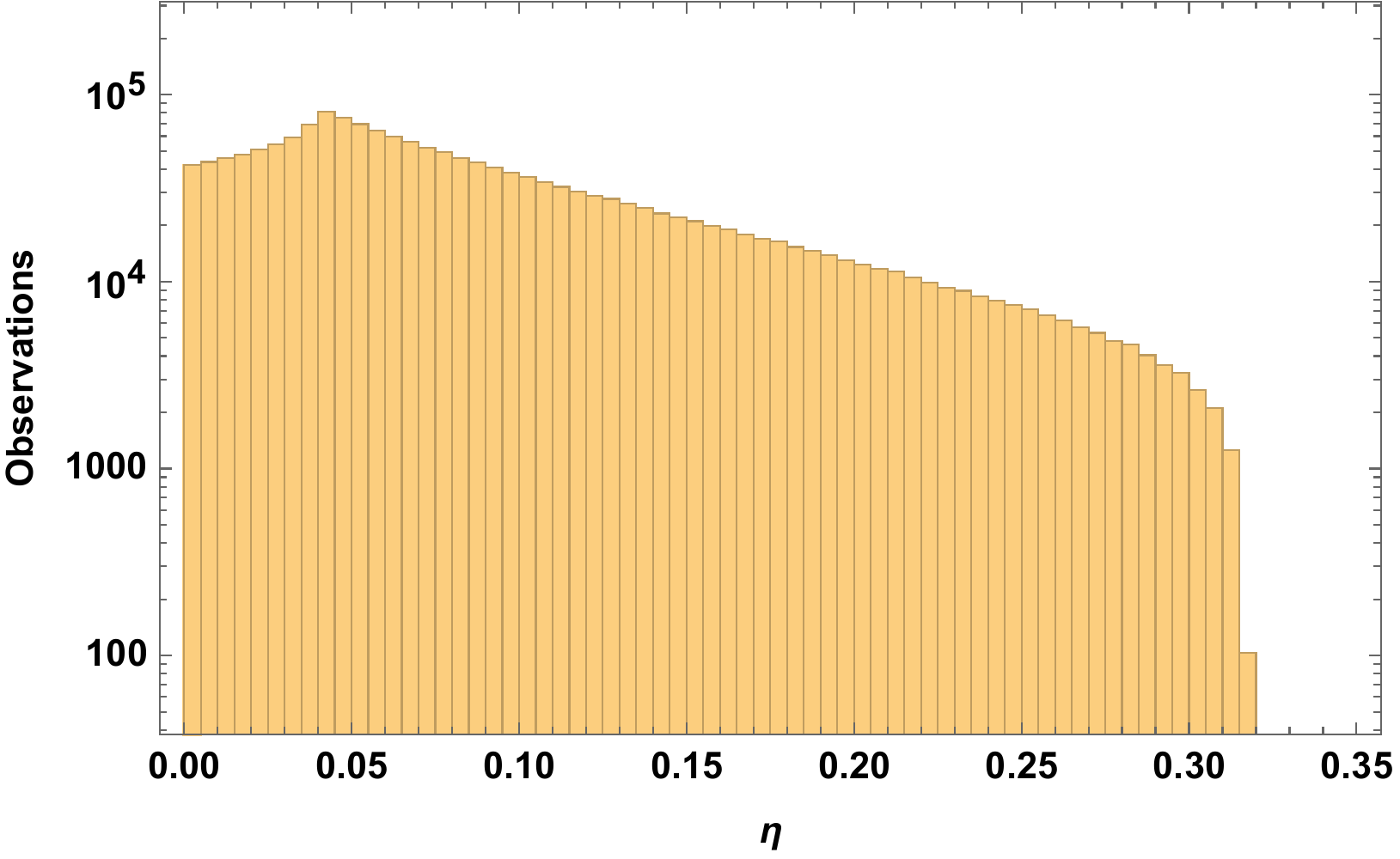}
 \end{center}
\caption{Histogram of sampled values of $\eta_{\rm SS}$ given in Eq. (\ref{effss}) for randomly sampling over a region of the parametric space ($\omega_c,\,\omega_h$).   The parameters are sampled over the uniform distributions $\omega_{c, h}\in [0,30]$ at fixed values of $\beta_c=1$ and $\beta_h=1/12$.  For plotting the histograms, we choose bin width of 0.01 to arrange $10^7$ data points. All the randomly generated values of $\eta_{\rm SS}$ lie below the efficiency $\eta^{\rm SS}_{\rm max}$ given in Eq. (\ref{etamaxss}), thus proving our point that $\eta^{\rm SS}_{\rm max}$ is upper bound on the efficiency in all operational regimes.}
\end{figure}
\subsection{Comparative  analysis of the engine at maximum work  to the engine at maximum efficient work}
In order to compare the performance of the engine operating at maximum work output to the engine operating at maximum efficient work function, we will     evaluate the fractional loss of work (denoted by  $R$), which is defined by the ratio of work lost (due to  entropy production) and the work extracted, in each case both for adiabatic and sudden switch driving protocols. The expression for $R$ is given by \cite{MePRR,Geva1994}
\begin{equation}
R = \frac{W_{\rm lost}}{W_{\rm ext}} = \frac{\eta_C}{\eta} - 1. \label{powerloss}
\end{equation}
First, we will discuss the adiabatic case. Using Eq. (\ref{EMEP}) in Eq. (\ref{powerloss}),  we obtain the following expression for the  ratio $R$ for the    engine operating in the maximum efficient work regime
\begin{equation}
R^{\rm EW}_{\rm AD}  =  \frac{1}{4} \left[\sqrt{(1-\eta_C)(9-\eta_C)}-(1-\eta_C)\right]. \label{R1}
\end{equation}
The corresponding expression for engine at maximum work can be obtained by substituting $\eta^{\rm W}_{\rm HT}=1-\sqrt{1-\eta_C}$ in Eq. (\ref{powerloss}), and we have
\begin{equation}
R^{\rm W}_{\rm AD}  =  \sqrt{1-\eta_C}. \label{R2}
\end{equation}
We plot Eqs. (\ref{R1}) and (\ref{R2}) in Fig. 4. We note that the dot-dashed blue curve  representing Eq. (\ref{R1}) lies well below the solid red curve representing Eq.(\ref{R2}), thus implying that the fractional loss of power at maximum efficient work function is much less than the corresponding case for the engine operating at maximum work output.  For  near-equilibrium conditions ($\eta_C\rightarrow 0$), we have $R^{\rm EW}_{\rm AD}/R^{\rm W}_{\rm AD}=1/2$, which implies that optimization of efficient work function leads to reduction in entropy production (or work lost) up to $50\%$ as compared to the optimization with respect to work output.

For the completeness sake, we  also plot the corresponding expressions for the fractional loss of power for the engine driven by sudden switch protocol. Since the relevant expressions are very complicated and not illuminating at all, we will not write them here and present numerical results only. In Fig. 4, dotted orange  and dashed brown curves represent the corresponding fractional loss of work for the engine working in the the maximum efficient work regime ($R^{\rm EW}_{\rm SS}$) and maximum work regime ($R^{\rm W}_{\rm SS}$), respectively.  Although the engine operating at maximum efficient work function wastes less work (power) as compared to the one at maximum work, the difference is very small.  Thus, in the sudden switch regime, the operation of the engine is dominated by the frictional effects and the choice of a trade-off objective function does not make much difference.

Finally, in the inset of Fig. 4, we plot the expressions for the ratio of the  extracted work at maximum efficient work function to the maximum work for adiabatic (solid gray curve) as well as sudden switch (dashed pink curve) case.   For the adiabatic (sudden switch) case, we find that the quantum harmonic Otto engine operating under the  condition of maximum efficient work produces at least $88.89\%$ ($99.93\%$) of the maximum work output. Considering the fact that in the adiabatic case, engine operating at maximum efficient work function can produce $88.89\%$ of the maximum work output while at the same time lowering the entropy production up to  $50\%$, efficient work function proves to be  a good objective function if our motive is to reduce the environment pollution and cut the fuel
consumption. As for the sudden switch case, we cannot make the same conclusion as explained below. The production of the $99.93\%$ of the maximum work output in efficient work regime does not imply that efficient work function is a good objective function to study engines operating in sudden-switch regime. To draw the full conclusion, we also have to consider the entropy production in maximum efficient work regime, which is not much different than that of in maximum work regime. This implies that  that in the sudden switch regime, maximum work and maximum efficient work function points lie very close to each other and optimal operation of the engine is almost  insensitive to the choice of  objective function.  In other words, in the sudden-switch regime,  the performance of the engine is dominated by the frictional effects which lead  to a large amount of entropy production,  thus rendering the choice of a trade-off objective function quite useless. 

\begin{figure}   
 \begin{center}
\includegraphics[width=8.6cm]{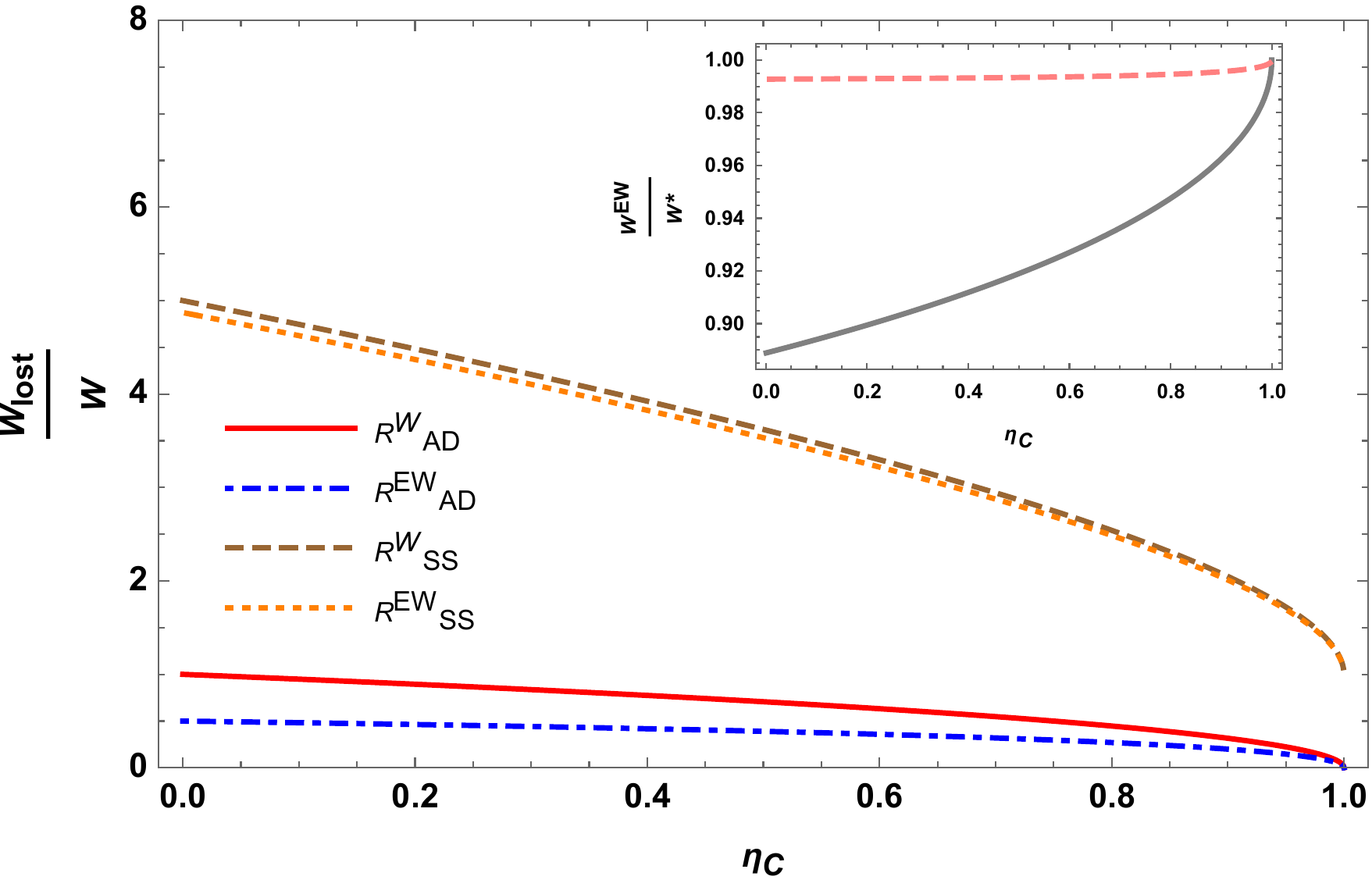}
 \end{center}
\caption{Comparison of the ratios of work lost due to entropy production  to the useful work output for two different optimization functions: efficient work function and work output. The lower-lying curves (solid red  and dot-dashed blue) represent the adiabatic case  whereas the upper lying curves (dashed brown and dotted orange) represent the corresponding case for the sudden-switch protocol. In the inset, we have plotted the expressions for the ratio of the  extracted work at maximum efficient work function to the maximum work for adiabatic (solid gray curve) as well as for the sudden switch case (dashed pink curve). }
\end{figure}
\section{Quantum Otto refrigerator}
In this section, we analyze the performance of the quantum harmonic Otto cycle working as a refrigerator for both the  adiabatic and nonadiabatic frequency modulations. For the refrigeration process,   work is invested to extract heat from the cold reservoir and dump it to the hot reservoir. Thus, we have $W_{\rm in}=- (Q_h+ Q_c)= W_1+ W_3>0$, $Q_c>0$ and $Q_h<0$. The coefficient of performance of the refrigerator is defined 
by
\begin{equation}
	\zeta = \frac{Q_c}{W_{\rm in}}=-\frac{Q_c}{Q_h+ Q_c}. 
	\label{cop}
\end{equation}
Using Eqs. (\ref{heat2}) and (\ref{heat4}) in Eq. (\ref{cop}), the COP takes the following form \cite{Abah2016EPL}:
\begin{equation}
	\zeta = \frac{\omega_c[\coth(\beta_c\omega_c/2) - \lambda\coth(\beta_c\omega_c/2)]}{(\lambda\omega_h-\omega_c)\coth(\beta_c\omega_c/2)-(\omega_h-\lambda\omega_c)\coth(\beta_h\omega_h/2)}. \label{cop22}
\end{equation}

In this section, we will investigate the performance analysis of the quantum harmonic Otto refrigerator by optimizing the so-called $\chi$-function, which is  defined by the product of the COP and cooling load of the refrigerator   \cite{YanChen1990}
\begin{equation}
	\chi = \zeta  \, {Q}_{c}.
\end{equation}
 As before,  we will discuss adiabatic case first and then the sudden-switch case. 
\subsection{Adiabatic driving}
 The optimal performance of the harmonic Otto refrigerator, undergoing adiabatic evolution, in the high-temperature regime was discussed in Ref. \cite{Abah2016EPL}. Here, we present the main results for the sake of completeness.
 Substituting $\lambda=1$ in Eq. (\ref{cop22}), the COP takes the following simple form
\begin{equation}
	\zeta_{\rm ad} = \frac{\omega_c}{\omega_h-\omega_c}=\frac{z}{1-z}, \label{cop2}
\end{equation}
where $z=\omega_c/\omega_h$. The COP at maximum $\chi$-function was found to be \cite{Abah2016EPL}
 \begin{equation}
 \zeta^{\chi}_{\rm HT} = \sqrt{1+\zeta_C}-1. \label{COPCA}
 \end{equation}
 As in the case of heat engines, above expression for the COP coincides with the COP of endoreversible \cite{YanChen1989} and symmetric low-dissipation \cite{deTomas2012} refrigerators, which is expected as we are working in the classical (high-temperature) limit. 
 
 Now, we will discuss the performance of the harmonic Otto refrigerator operating under the conditions of maximum $\chi$-function in the low-temperature regime. This case has not been discussed in the literature yet. In the low-temperature limit, the $\chi$-function reads as follows
 \begin{equation}
 \chi = \frac{\omega_c^2}{\omega_h-\omega_c} \left( e^{-\beta_c \omega_c} -e^{-\beta_h \omega_h}     \right). \label{chiadiabatic}
 \end{equation}
 Optimizing  Eq. (\ref{chiadiabatic}) with respect to $\omega_h$ and $\omega_c$, i. e., setting $\partial \chi/\partial \omega_h=0$ and $\partial \chi/\partial \omega_c=0$, we obtain following two equations, respectively 
 \begin{eqnarray}
 e^{\beta_h \omega_h-\beta_c \omega_c} &=&   = 1 + \frac{\beta_c \omega_c \zeta_C}{\zeta (1+\zeta_C)}, \label{XX1}
 \\
 e^{\beta_h \omega_h-\beta_c \omega_c} &=&    1 +  \frac{\beta_c \omega_c}{2+\zeta-\beta_c \omega_c}, \label{XX2}
 \end{eqnarray}
 where we have used the relations $\zeta=\omega_c/(\omega_h-\omega_c)$ and $\zeta_C=\beta_h/(\beta_h-\beta_c)$. These equations cannot be solved analytically for $\omega_h$ and $\omega_c$. However,   Eqs. (\ref{XX1}) and (\ref{XX2}) can be combined to yield the following transcendental equation  
 \begin{equation}
 \frac{(2\zeta_C-\zeta)(\zeta_C-\zeta)}{\zeta(1+\zeta)\zeta_C} = \ln \left[\frac{(2+\zeta)\zeta_C} {\zeta(1+\zeta_C)}  \right]  , \label{trans}
 \end{equation}
which implies that COP under the conditions of maximum $\chi$-function depends on  $\zeta_C$ only and is independent of the parameters of the system. Eq.  (\ref{trans}) (dashed red curve labelled by $\zeta^{\chi}_{\rm LT}$) is plotted in Fig. 3 along with the expression for the COP given in Eq. (\ref{COPCA}) (solid blure curve) in the high-temperature limit. It is evident from Fig. 3 that the COP of the harmonic Otto refrigerator in the low-temperature regime is higher than that of operating in the high-temperature regime, although the difference is very small.

\subsection{Sudden switch of frequencies}

Now, we are ready to   discuss the case in which frequency of the harmonic oscillator is switched suddenly from one value to the other. In this case, $\lambda=(\omega_c^2+\omega_h^2)/2 \omega_c \omega_h$. In the sudden switch regime, the cooling load can only be be maximized for the vanishing COP, which is not a useful result. Hence, to study the optimal operation of the refrigerator, $\chi$-function presents us with a sensible choice.  For sudden-switch driving protocol, in the high-temperature limit, the expression for the cooling load and input work are evaluated to be, %
\begin{equation}
Q^{\rm SS}_c=\frac{1}{\beta_h}\left[\tau -\frac{1}{2} \left(z^2+1\right)\right], \quad W^{\rm SS}_{\rm in}= \frac{\left(z^2-1\right) \left(z^2-\tau \right)}{2 \beta_h z^2}.
\label{coolingSS}
\end{equation}
Further, the COP, $\zeta=Q_c/W_{\rm iin}$, takes the form
\begin{equation}
\zeta_{\rm SS} = \frac{z^2 \left(2 \tau -z^2-1\right)}{\left(z^2-1\right) \left(z^2-\tau \right)} \label{copss}.
\end{equation}
Before proceeding further, we would like to make some comments about the performance of harmonic Otto refrigerator  operating in the sudden-switch limit.   First, we can see from Eq. (\ref{coolingSS}) that the positive cooling condition ($Q^{\rm SS}_c>0$)  lead to the following constraint on the refrigeration process: 
\begin{equation}
 z^2+1<2\tau               . \label{constraint}
 \end{equation}
Eq. (\ref{constraint}) puts more stringent condition on the cooling condition for a nonadiabatic harmonic Otto refrigerator  as compared to its adiabatic counterpart in which positive cooling condition is  given by $z<\tau$. Thus, for the given ratio  reservoir temperatures, the compression ratio $z$ for the sudden switch (in general for any nonadiabatic driving protocol) case should be large as compared as its adiabatic counterpart to extract heat from the cold reservoir in a refrigeration cycle. Further, Eq. (\ref{constraint}) puts another constraint on the reservoir temperatures \cite{VOzgur2020}. The cooling condition $ z^2+1<2\tau$ implies that $z<\sqrt{2\tau-1}$. This constraint is meaningful only  when $\tau>1/2$, which in turn implies that the quantum harmonic Otto refrigerator    cannot extract heat from the cold reservoir   unless the temperature of the cold reservoir is greater than half   the temperature of the hot reservoir. This is due to the generation of the friction in the adiabatic (in the thermodynamic sense) branches when we drive our oscillator nonadiabatically \cite{VOzgur2020,Plastina2014}. For more information, we refer our readers to the Ref. \cite{VOzgur2020}.

 \begin{figure}    [H]
 \begin{center}
\includegraphics[width=8.6cm]{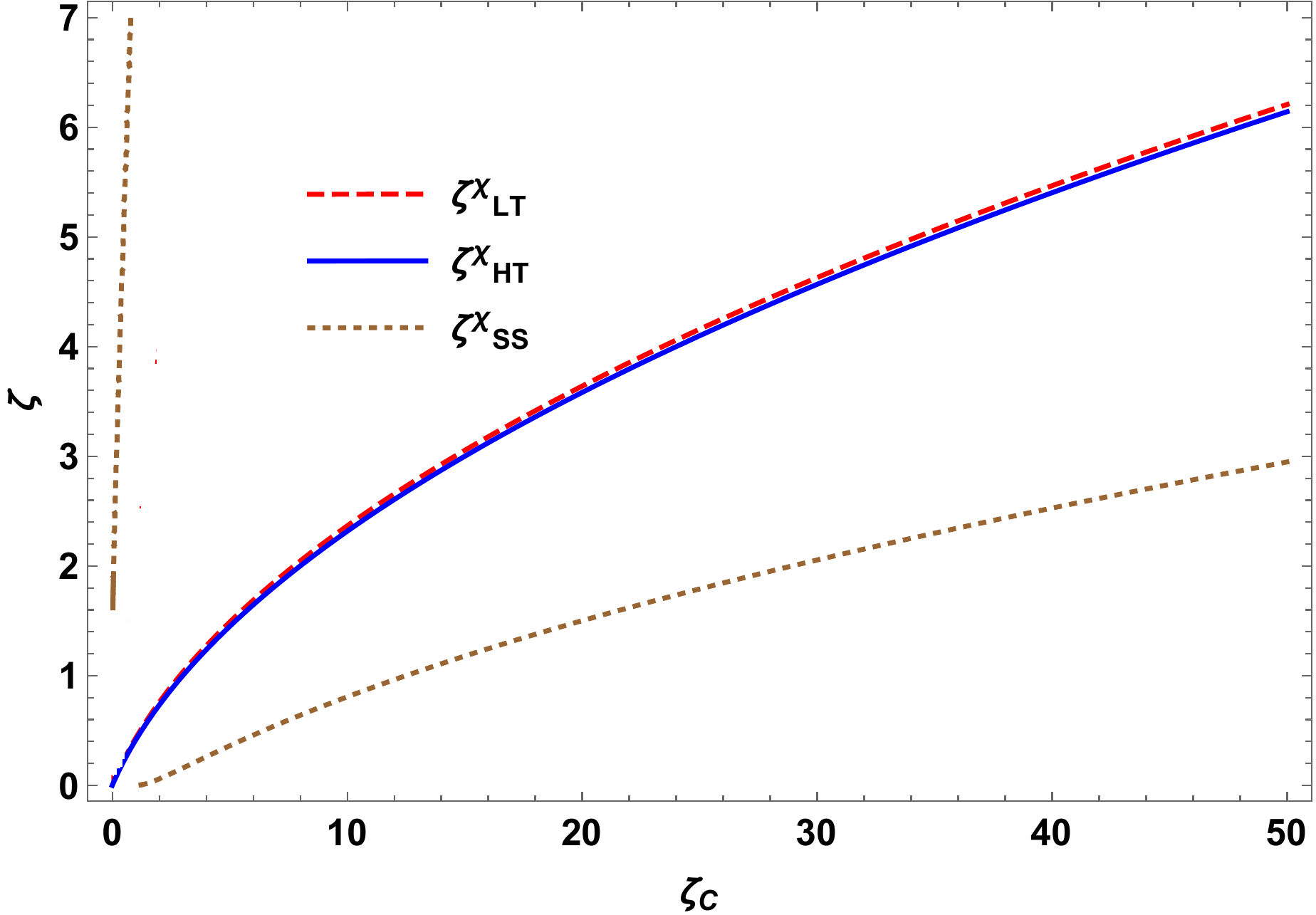}
 \end{center}
\caption{COP at maximum  $\chi$-function versus Carnot COP  $\zeta_C$. For the adiabatic case, solid blue curve represents the low-temperature limit, Eq. (\ref{trans}), while the solid red curve represents  the high-temperature limit, Eq. (\ref{COPCA}). Solid brown curve represent the sudden-switch case, Eq. (\ref{zetachiss}).}
\end{figure}
\subsubsection{Upper bound on the COP}
Now, we will show that in the sudden switch regime, the maximum COP of the engine is not given by the Carnot COP, $\zeta_C=\beta_h/(\beta_c-\beta_h)$. Optimizing Eq. (\ref{copss}) with respect to $z$ and substituting the resulting expression for $z$ in Eq. (\ref{copss}), we obtain the following expression for maximum COP in the sudden-switch regime,
\begin{equation}
\zeta_{\rm max} = 1 + 3\zeta_C -2\sqrt{2\zeta_C(1+\zeta_C)}. \label{copmax}
\end{equation} 

The same expression for the maximum COP was also derived in a recent paper  \cite{VOzgur2020}.  The method used in the present paper is much simpler than the method used in Ref. \cite{VOzgur2020}. We also notice that  $\zeta_{\rm max}$  given in Eq. (\ref{copmax}) is  much smaller than the Carnot COP.  Similar to the case of heat engine, the expression given in Eq. (\ref{copmax}) is an valid upper bound on the efficiency  for all operational regime in spite of the fact that Eq. (\ref{copmax}) is derived in high-temperature regime. We checked this numerically by plotting a histogram  (not shown here) similar to Fig. 3.

\subsubsection{Optimization of $\chi$-function}
Now we turn to the optimization of the $\chi$-function, whose expression can be found by just multiplying $Q_c$ given in  Eq. (\ref{coolingSS}) with $\zeta_{\rm SS}$  (Eq. (\ref{copss})). Optimization of the  resulting equation with respect to parameter $z$ yields the following equation:
\begin{equation}
y^3-3 y^2+3 \tau  y+\tau  (1-2 \tau )=0, \label{cubiceq}
\end{equation}
where $y=z^2$. Due to casus irreducibilis (see Appendix A), the roots of the above cubic equation can only be expressed using complex radicals, although the roots are actually real \cite{BarnettBook} . Still, we can plot  the efficiency as a function of Carnot efficiency $\eta_C$ by employing the following method. Using Eq. (\ref{copss}), we can express $z$ in terms of terms of $\zeta$ and $\tau$:
\begin{widetext}
\begin{equation}
z^2=\frac{\zeta  \tau -\sqrt{((\zeta +2) \tau +\zeta -1)^2-4 \zeta  (\zeta +1) \tau }+\zeta +2 \tau -1}{2 (\zeta +1)}.
\end{equation}
Using this expression for $z$ in Eq. (\ref{coolingSS}) and optimization of the resulting expression for $\chi=\zeta \, Q_c$ with respect to $\zeta$ yields the following equation,
\begin{equation}
\zeta  (\zeta +1) (\tau -1) [3F+\zeta  (\tau -1)+2 \tau +1]+F [F+3 \zeta  (\tau -1)+2 \tau -1]=0, \label{zetachiss}
\end{equation}
\end{widetext}
where $F=\sqrt{((\zeta +2) \tau +\zeta -1)^2-4 \zeta  (\zeta +1) \tau }$, which shows that COP $\zeta$, at optimal $\chi$-function, depends  on ratio of reservoir temperatures ($\tau$) only. We have plotted Eq. (\ref{zetachiss}) in Fig. 3 (lowest lying dotted brown curve labelled as $\zeta^{\chi}_{\rm SS}$). It is self evident from Fig. 3 that  refrigerator operating in sudden switch regime is far less efficient than than its adiabatic counterpart, which is expected result as nonadiabatic driving induces frictional effects in the operation of the engine.

Unlike the case of heat engine, we cannot carry out  a comparative analysis for the refrigerator as the COP at maximum cooling load vanishes and this is not very encouraging result to have comparison with.

 \section{Conclusions}
We have analyzed the optimal performance of a harmonic quantum Otto cycle working under the conditions of a maximum trade-off figure of merit, which pays equal attention to the efficiency (COP) and the work output (cooling load) of the engine (refrigerator). In the case of heat engine, the chosen trade-off objective function is the efficient work function while the refrigerator is studied under the optimal conditions of maximum $\chi$-function. First,   we obtained the analytic expressions for the efficiency of the adiabatically (slowly) driven  engine in the    high-temperature   and low-temperature regimes.   Further, in the sudden switch regime, first we  obtained the analytic expression for the upper bound on the efficiency 
and then we derive the analytic expression for the efficiency under the optimal conditions. Furthermore, by carrying out a comparative analysis between the engines operating at maximum efficient work function and  maximum work, we  showed that the efficient work function is a   good objective function only for the adiabatic driving protocol whereas for the sudden-switch protocol  the choice of the objective function does not make much difference. 
We repeated our analysis to investigate  the optimal performance of the harmonic Otto refrigerator  and obtained analytic results under various operational regimes. Finally, we would like to add that in our study, we have obtained analytic expressions for  the efficiency and the COP for two extreme driving protocols: adiabatic and sudden switch.  The actual performance of of thermal machine under consideration will lie in between these two regimes. We hope that the detailed analysis presented here will add a valuable reference to the existing literature on quantum heat engines.  

\section{Acknowledgements} 
Varinder Singh acknowledges support by the Institute for Basic Science in Korea (IBS-R024-D1).  
\appendix
\section{CASUS IRREDUCIBILIS}
In cubic equations, the case of Casus irreducibilis may arise \cite{Kleiner2007,Stewart1990} when the discriminant $D=18abcd-4b^3d+b^2c^2-4ac^3-27a^2d^2$ of the equation 
\begin{equation}
a x^3 + b x^2 + c x + d  =0, \qquad (a, b, c, d \text{\, are real})
\end{equation}
is  positive, i.e., $D>0$.
In such a case, all three roots of the cubic equation are real and distinct.   If the rational root test cannot be used to find the roots, 
then the given polynomial is Casus irreducibilis and complex valued expressions (expressions containing complex numbers) are needed to 
express the roots in radicals.

In our case,  we have the following cubic equation (see Eq. (\ref{cubiceq}))
\begin{equation}
y^3-3 y^2+3 \tau  y+\tau  (1-2 \tau )=0, \label{cubiceq2}
\end{equation}
The discriminant $D$ of  Eq. (\ref{cubiceq2}) is given by
\begin{equation}
D = 108(1-\tau)^3\tau
\end{equation}
As $0<\tau<1$, $D>0$.  Hence, we are dealing with a cubic equation (Eq. (\ref{cubiceq})) which presents us with the case of Casus irreducibilis.
\label{casusir}

\bibliography{QHEbiblo.bib}
\bibliographystyle{apsrev4-2}

\end{document}